\documentclass[aps,prl,twocolumn,superscriptaddress,raggedbottom,showpacs,floatfix]{revtex4-1}
\usepackage{color}
\usepackage{graphicx,latexsym}
\usepackage{epstopdf}
\usepackage{comment,bm}

\usepackage{amsmath}
\usepackage{amssymb}
\usepackage{mathtools}
\usepackage{graphics}
\usepackage{comment}
\usepackage{tikz}

\newcommand{\beq}{\begin{equation}}
\newcommand{\eeq}{\end{equation}}
\newcommand{\bei}{\begin{itemize}}
\newcommand{\eei}{\end{itemize}}
\newcommand{\ben}{\begin{enumerate}}
\newcommand{\een}{\end{enumerate}}

\newcommand{\be}{{\mathbf e}}

\usepackage{hyperref}
\hypersetup{
   pdfpagemode=None, 
   pdfstartpage=1,
   pdfmenubar=true,
   pdftoolbar=true,
   colorlinks = true,
   linkcolor=blue,
   citecolor=blue,
   urlcolor=blue,
   bookmarksopen=false
 }

\definecolor{darkblue}{rgb}{0.,0.24,0.51}
\definecolor{britishracinggreen}{rgb}{0.0, 0.26, 0.15}
\definecolor{darkgreen}{rgb}{0,0.60,.2}

\def\be{\begin{equation}}
\def\ee{\end{equation}}

\begin{document}
\renewcommand{\vec}{\mathbf}
\renewcommand{\Re}{\mathop{\mathrm{Re}}\nolimits}
\renewcommand{\Im}{\mathop{\mathrm{Im}}\nolimits}

\title{Hybrid dark excitons in monolayer $\hbox{MoS}_2$}

\author{Hong Liu$^{1,2}$, Anny Pau$^{1}$, and Dmitry K. Efimkin}
\affiliation{School of Physics and Astronomy, Monash University, Victoria 3800, Australia}
\affiliation{ARC Centre of Excellence in Future Low-Energy Electronics Technologies, Monash University, Victoria 3800, Australia}

\begin{abstract}
Transition metal dichalcogenides have a rich exciton landscape consisting of a variety of bright and dark excitonic states. We consider the lowest-energy dark states in $\hbox{MoS}_2$, which can be referred to as hybrid excitons, as they are formed by a Dirac electron and a Schr\"{o}dinger hole. The chiral nature of the Dirac electron introduces asymmetry to the excited exciton state spectrum and couples the relative motion of the electron and hole with the center-of-mass motion. We demonstrate that this coupling generates an additional contribution to the Berry curvature of hybrid excitons. The additional contribution is substrate-dependent and accounts for almost one quarter of the Berry curvature in suspended samples. The nontrivial geometry of hybrid excitons is manifested in the optical anomalous valley Hall effect, which can be observed via recently developed pump-probe photoemission spectroscopy.  We argue that the Hall angle of hybrid excitons is approximately one half of that for bright excitons. Moreover, the anticipated long lifetime of hybrid excitons favors an extended propagation distance and allows the spatial separation of hybrid excitons with different valley indices. 
\end{abstract}

\date{\today}
\maketitle
\section{I. Introduction}
Monolayer transition metal
dichalcogenides (TMDs), such as $\hbox{MoS}_2$, $\hbox{MoSe}_2$, $\hbox{WS}_2$, and $\hbox{WSe}_2$, have recently attracted substantial interest due to their unique optical and electronic properties~\cite{ReviewGeneral}. The optical response of TMDs is governed by bright excitons, which have exceptionally large binding energies and exhibit strong coupling with light. Due to the presence of a valley degree of freedom and valley selection via circularly polarized light, TMD monolayers and heterostructures have great potential for optoelectronics and valleytronics~\cite{ExcitonReview1,ExcitonReview2,ExcitonReviewGlazov,ReviewValleytronics1,ReviewValleytronics2,ReviewValleytronics3}. 

Another unique feature of bright excitons is their geometrical or Berry phase~\cite{XBerryPhaseExchange1,XBerryPhaseExchange2,ExcitonBerryPhaseMacDonald,ExcitonBerryPhasePolariton,XBerryPhaseTrushin}, which is inherited from the nontrivial band geometry of electrons and holes in TMDs~\cite{BC-RMP}. The nonzero valley-dependent Berry curvature alters the exciton dynamics in a manner similar to that of an effective magnetic field in momentum space. For instance, this curvature results in a deflection of photoexcited excitons, which is usually referred as the optical anomalous valley Hall effect (AVHE). This effect allows the spatial separation of excitons from two valleys. The optical AVHE has been observed in TMD monolayers and heterostructures~\cite{EHE-2017,EHE-2020} and can be enhanced in the presence of an optical cavity~\cite{EHE-2019}. However, further progress has been hampered by the very short (sub-picosecond) valley coherence time, which is limited by strong intervalley exchange interactions~\cite{IntervalleyExchange1,IntervalleyExchange3,IntervalleyExchange4}. 

The electronic band structure of TMD monolayers demonstrates strong spin–orbit coupling and includes multiple conduction minima/valence band maxima. Recent theoretical studies have shown that in addition to bright exciton states, a variety of dark species also exist~\cite{MalicMain}. These states are momentum- and/or spin-forbidden because photons cannot provide the required momentum or flip spin to induce an interband transition to these states. Despite being optically inactive, the dark states play a significant role in nonequilibrium dynamics; for example, the dark states determine the efficiency of TMD light emission. Importantly, dark excitons can accumulate if their energies are below the energies of bright excitons, as predicted for $\hbox{MoS}_2$, $\hbox{WS}_2$, and $\hbox{WSe}_2$~\cite{MalicMain}. The rich landscape of dark excitons is still relatively unexplored (one exception is momentum-allowed but spin-forbidden states that can be brightened by a magnetic field~\cite{SpinForbidden1,SpinForbidden2,SpinForbidden4} or by coupling with surface plasmon-polaritons~\cite{SpinForbidden3}) due to a lack of effective approaches for directly probing dark states~\cite{DarkExcitonsMapping1,DarkExcitonsMapping2,DarkExcitonsMapping3}. This barrier has recently been lifted by the development of a pump-probe photoemission technique that can reveal the spatial, temporal, and spectral dynamics of both bright and dark excitons~\cite{ExpImportant}, which has opened a new area of dark excitonics in TMD monolayers and beyond. 

\begin{figure}[t]
\vspace{-0.3cm}
	\begin{center}
		\includegraphics[trim=2cm 11cm 15cm 4.1cm, clip, width=1.0\columnwidth]{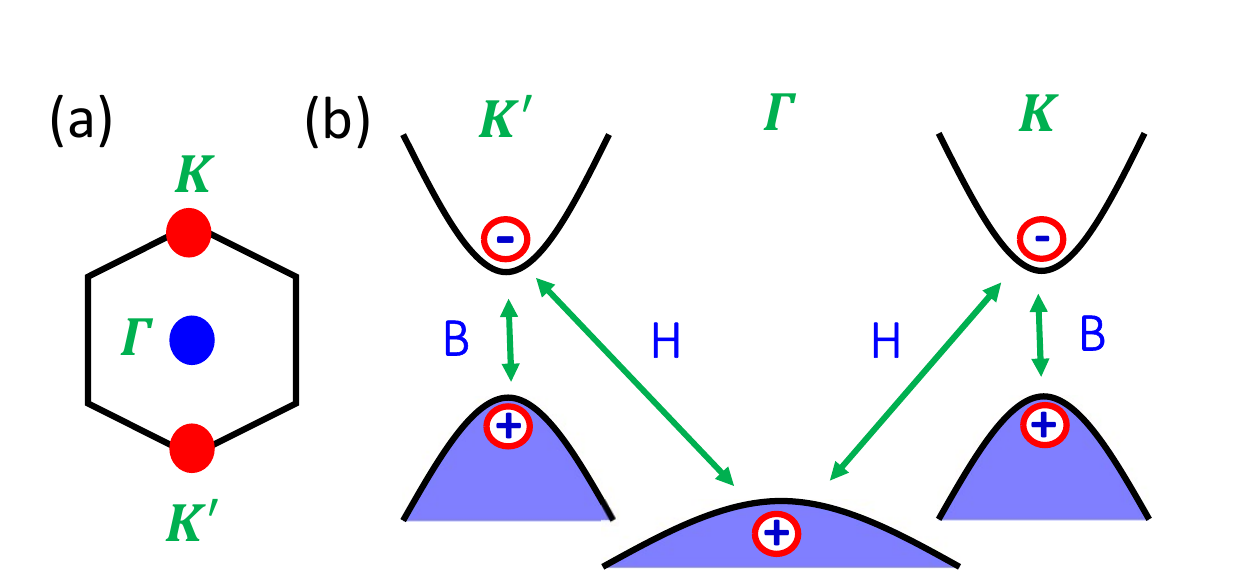}
		\caption{(a) Brillouin zone of an $\hbox{MoS}_2$ monolayer. (b) Low-energy band structure with dark hybrid dark (H) and bright (B) excitonic states. While single-particle direct gap at $K$ ($K'$) valley is smaller than the indirect one involving holes near the $\Gamma$-point, hybrid dark excitons formed by light Dirac electrons and heavy Schr\"{o}dinger holes have the lowest energy }
		\label{Fig1}
	\end{center}
	\vspace{-0.4cm}
\end{figure}

Here, we focus on $\hbox{MoS}_2$, for which the lowest-energy excitonic states have been predicted~\cite{MalicMain} to arise from a Dirac electron in the vicinity of one of two nonequivalent valleys and a Schr\"{o}dinger hole in the vicinity of $\Gamma$-point. These states are sketched in Fig.~\ref{Fig1}. Due to the mixing of dispersion relations of different natures, we refer to these states as hybrid excitons~\footnote{It should be noted that the asymmetry in the dispersion relations for an electron and a hole forming an exciton is common in semiconductors. For instance, the dispersion of electrons in III-V semiconductors as well as in Si and Ge can be approximated to be parabolic, but the dispersion of holes is strongly effected by spin-orbit interactions~\cite{Lipari1,Lipari2}}. Recent photoluminescence experiments~\cite{HybridExciotnsExp} supported by DFT calculations have demonstrated that hybrid excitons effectively accumulate after bright exciton photoexcitation in hBN-encapsulated monolayer $\hbox{MoS}_2$, with an energy $83~\hbox{meV}$ below the bright exciton energy. In the present work, we incorporate the previously neglected Dirac nature of electrons participating in hybrid excitons and uncover the nontrivial geometry of hybrid excitons and its manifestations in the optical AVHE. 

We argue that the chiral nature of Dirac electrons introduces asymmetry in the hybrid exciton excited state spectrum and couples the relative motion of the electron and hole with the center-of-mass motion. We demonstrate that this coupling generates an additional contribution to the Berry curvature of the exciton that is substrate-dependent and accounts for almost one quarter of the Berry curvature in suspended samples. We analyze the intrinsic optical AVHE mediated by hybrid excitons and argue that their Hall angle is approximately one half of that for bright excitons. The long lifetime for hybrid dark excitons favors their long-distance propagation, which can be tracked by a recently developed photoemission technique~\cite{ExpImportant}.

The remainder of this paper is organized as follows. Section~II introduces our model describing the low-energy band structure of $\hbox{MoS}_2$. Section~III is devoted to the spectrum of hybrid excitons at rest. Section~IV presents the coupling between the center-of-mass motion and the relative motion for the electron and hole.  Section~V introduces the Berry curvature of excitons and identifies three contributions to this curvature, which arise from different physical origins. Section~VI provides an analysis of the optical AVHE mediated by dark excitons. A discussion and conclusions are given in Section~VII. 

\section{II. Model}
The low-energy electronic structure of the $\hbox{MoS}_2$ monolayer  that is sufficient to describe hybrid excitons is presented in Fig.~\ref{Fig1}. Electronic states reside near $K$ ($\alpha=1$) or $K'$ ($\alpha=-1$) valleys, which have opposite spin directions. Hole states reside near the $\Gamma$-point and have the usual spin degeneracy. The kinetic energy of electron and hole states can be described by the following Hamiltonian 
\begin{equation}
H_0=\sum_{\vec{p}} \left( \varepsilon^\mathrm{c}_\vec{p}c^\dagger_{\vec{p}} c_{\vec{p}} + \varepsilon^\mathrm{v}_\vec{p} v^\dagger_{\vec{p}} v_\vec{p} \right).     
\end{equation}
Here, $\varepsilon^\mathrm{v}_\vec{p}=-\vec{p}^2/2m_\mathrm{v}$ is the conventional dispersion relation for valence band electrons, and $m_\mathrm{v}$ is their mass. The dispersion of Dirac electrons is $\varepsilon_{\vec{p}}^\mathrm{c}= \sqrt{(v_\mathrm{D} p)^2+\Delta^2}$ with velocity $v_\mathrm{D}$ and direct bandgap $2 |\Delta|$. The behavior at the bottom of the band is also quadratic as $\varepsilon^\mathrm{c}_\vec{p}=\vec{p^2}/2m_\mathrm{c}$ with mass $m_\mathrm{c}=\Delta/v^2$ while the valley-dependent chirality of Dirac electrons is encoded in their spinor wave function   
\begin{equation}
\label{DiracSpinor}
    |\vec{p}\rangle_\mathrm{c}=\left(
\begin{array}{c}
\alpha \cos\frac{\theta_{\vec p}}{2} \\
\sin\frac{\theta_{\vec p}}{2}e^{i\alpha\phi_{\vec p}}
\end{array}
\right).
\end{equation}
Here,  $\phi_{\vec p}$ is the polar angle for the momentum $\vec{p}$, and $\cos\theta_{\vec{p}}=\Delta/\varepsilon_\vec{p}$. It is instructive to introduce the compact notations $c_{\vec{p}}\equiv\cos(\theta_{\vec p}/2)$ and $s_{\vec{p}}=\sin(\theta_{\vec p}/2)$. The important feature of Dirac electrons is their geometrically nontrivial spectrum. The geometry is characterized by the Berry connection $\vec{A}^\mathrm{c}_{\vec{p}}=i\langle\vec{p}|\nabla_\vec{p}|\vec{p}\rangle_\mathrm{c}$ and the Berry curvature $\Omega^\mathrm{c}_{\vec{p}}= [{\bm \nabla_\vec{p}}\times {\bm A}^\mathrm{c}_{\vec{p}}]_\mathrm{z}$. The latter is given by~\cite{BerryPhaseCulcer} 
\begin{equation}
\label{ElectronBC}
{\bm \Omega}_{\vec{p}}^{\mathrm{c}}=-\frac{\alpha v_\mathrm{D}^2 \Delta}{2(v_\mathrm{D}^2 \vec{p}^2+\Delta^2)^{3/2}}.
\end{equation}
The sign of the Berry curvature is valley-dependent, while its amplitude smoothly decreases at the momentum scale $p_\mathrm{\Delta}=\Delta/v_\mathrm{D}$, where the dispersion of the Dirac electron evolves from quadratic behavior to linear one. We will demonstrate below that the nontrivial band geometry of the electrons is passed on to the hybrid excitons. The latter are formed by attractive Coulomb interactions between electrons and holes, which can be described by the following Hamiltonian
\begin{equation}
H_\text{C}=\sum_{\vec{p}\vec{p}'\vec{q}} \Lambda_{\vec{p}+\vec{q},\vec{p}} V_\vec{q}\;  c^\dagger_{\vec{p}+\vec{q}}  v^\dagger _{\vec{p}'-\vec{q}}  v_{\vec{p'}} c_{\vec{p}}. 
\end{equation}
Here, the angle factor $\Lambda_{\vec{p}+\vec{q},\vec{p}}=\langle \vec{p}+\vec{q}|\vec{p}\rangle$ is given by the overlap of spinor wave functions for Dirac electrons. The Keldysh potential $V_q=2\pi e^2 F(q)/\epsilon q$ with $F(q)=1/(1+r_0 q)$ incorporates dielectric screening in TMD monolayers deposited on a dielectric substrate~\cite{PhysRevB.84.085406,srep39844} and has been argued to accurately describe excitonic states~\cite{PhysRevB.88.045318,PhysRevLett.113.076802,ExcitonBerryPhaseMacDonald}. The screening length $r_0 =\chi_{\mathrm{2D}}/(2\epsilon)$ is given by the 2D polarizability $\chi_{\mathrm{2D}}$ of the TMD layer. The dielectric constant is $\epsilon=(\epsilon_\text{T}+\epsilon_\text{B})/2$, with $\epsilon_\mathrm{T}$ and $\epsilon_\mathrm{B}$ as dielectric constants for the environment above and below the TMD monolayer. 

For the numerical calculations presented below, we will use the following set of parameters for the $\hbox{MoS}_2$ monolayer: $\Delta\approx0.79~\hbox{eV}$, $v_\mathrm{D}=0.53\cdot10^6\; \hbox{m}/\hbox{s}$, and $\chi_{\mathrm{2D}}\approx68\; \hbox{\AA}$~\cite{ExcitonBerryPhaseMacDonald,ExcFarhan,Korm_nyos_2015}. The resulting mass of Dirac electrons $m_\mathrm{c}\approx 0.5\; m_0$ is one fifth of the mass of holes $m_\mathrm{v}\approx 2.5\; m_0$, where $m_0$ is the free electron mass. Unless otherwise stated, we will assume that $\hbox{MoS}_2$ lies on an $\hbox{SiO}_2$  substrate ($\epsilon_\mathrm{B}=3.9$) and is exposed to air ($\epsilon_\mathrm{T}=1$).

\begin{figure}[t]
	\begin{center}
		\includegraphics[trim=0cm 0cm 0cm 0cm, clip, width=0.6\columnwidth]{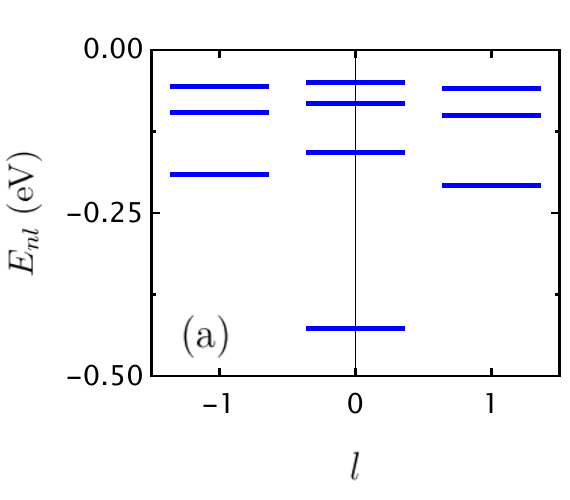}
		\includegraphics[trim=5cm 8.1cm 19cm 5.5cm, clip, width=0.37\columnwidth]{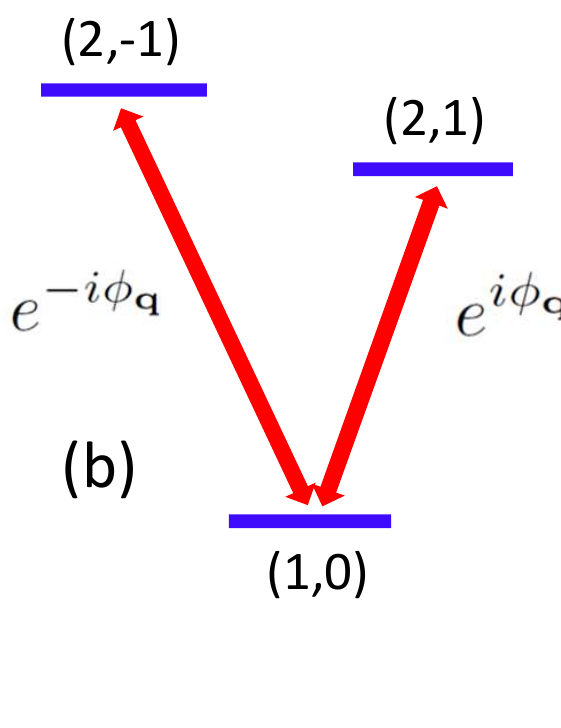}
		\caption{(a) Spectrum of $\Gamma K'$ hybrid excitons with orbital quantum number $l=0, \; \pm 1$ (see also Table~\ref{Tab1}). The chiral nature of Dirac electrons results in spectrum asymmetry $E_{n,+}\neq E_{n,-}$. The spectrum of $\Gamma K'$ can be obtained by mirror inversion.  (b) This sketch illustrates the coupling between ground and excited states induced by exciton motion. The corresponding matrix elements include the phase winding factors $e^{\pm i \phi_\vec{q}}$ and result in an additional contribution to the Berry curvature of the hybrid excitons (see Eq.~(\ref{BerryCurvatureFull})).}
		\label{Fig2}
	\end{center}
\end{figure}

\section{III. Hybrid excitons}

An exciton state represents a hydrogen-like bound state formed by an electron and a hole and is given by 
\begin{equation}
\label{ExcitonGeneral}
|\vec{q}\rangle_\mathrm{X}=\sum_{\vec{p}} C_{\vec{p}\vec{q}} c^\dagger_{\vec{p}_+} v_{\vec{p}_-}, \quad \quad \vec{p}_\pm=\vec{p}\pm\beta_{\mathrm{c}(\mathrm{v})} \vec{q}.
\end{equation}
Here, $\vec{p}$ is the relative momentum of the electron-hole pair, and $\vec{q}$ is their center-of-mass momentum. The factors $\beta_\mathrm{c}=m_\mathrm{c}/M$ and $\beta_\mathrm{v}=m_\mathrm{v}/M$ with $M=m_\mathrm{c}+m_\mathrm{v}$ determine how the total momentum $\vec{q}$ is redistributed between the electron and hole. The envelope wave function in momentum space $C_\vec{pq}$ satisfies the following eigenvalue problem
\begin{equation}
\label{EigenvalueProblemZeroQ}
(\varepsilon^\mathrm{c}_{\vec{p}_+}-\varepsilon^\mathrm{v}_{\vec{p}_-}) C_{\vec{p}\vec{q}} - \sum_{\vec{p}'} V_{\vec{p}-\vec{p}'}\Lambda_{\vec{p}_+ \vec{p}'_+} C_{\vec{p}'}= E_\vec{q} C_{\vec{p}\vec{q}},
\end{equation}
where $E_\vec{q}$ is the energy of the exciton. The chiral nature of Dirac electrons is reflected via the angle factor in the integral given by 
\begin{equation}
\label{AngleFactor}
\Lambda_{\vec{p}_+ \vec{p}'_+}=c_{\vec{p}_+} c_{\vec{p}'_+}  + s_{\vec{p}_+} s_{\vec{p}'_+} e^{-i \alpha (\phi_{\vec{p}_+}- \phi_{\vec{p}'_+})}.  
\end{equation}
Here $c_\vec{p}=\cos(\theta_\vec{p}/2)$ and $s_\vec{p}=\sin(\theta_\vec{p}/2)$ are the factors, which have been introduced below Eq.~(\ref{DiracSpinor}). The presence of the angle factor $\Lambda_{\vec{p}_+ \vec{p}'_+}$ strongly impacts the spectrum of states for a hybrid exciton at rest ($\vec{q}=0$), as presented in Fig.~\ref{Fig2} and Table~\ref{Tab1}. The states are labeled by the main $n=1,2,...$ and orbital $l=0,\pm1,...$ quantum numbers. Due to the heavy mass of the hole, the binding energy of the ground state $E^\mathrm{H}_{10}\approx 0.43~\hbox{eV}$ is larger than that for bright excitons $E_{10}^{\mathrm{B}}\approx 0.34~\hbox{eV}$, which has also been evaluated and shown to agree with previous calculations~\cite{ExcitonBerryPhaseMacDonald,ExcFarhan}. The spectrum of excited states is non-hydrogenic (does not follow $(n-1/2)^{-2}$ behavior) and exhibits asymmetry between excited states with opposite angular momentum quantum number $E_{n,+}\neq E_{n,-}$. This asymmetry is a signature of nontrivial band geometry for electrons and/or holes and appears not only in TMD monolayers~\cite{BerryExciton1,BerryExciton2} but also in other physical systems~\cite{OtherExcioton1,OtherExcioton2,OtherExcioton3}. The splitting $\Delta E_{2,\pm 1}\approx16~\hbox{meV}$ is largest between the first excited states with $n=2$ and quickly decreases with the main quantum number $n$. As we demonstrate below, this asymmetry is an essential component of one of the additional contributions to the Berry curvature of hybrid excitons.

\begin{table}
\begin{center}
\begin{tabular}{|p{1.4cm}|p{1.4cm}|p{1.4cm}|p{1.4cm}|p{1.4cm}|}
\hline
  & $n=1$ & $n=2$ & $n=3$ & $n=4$ \\
 \hline
 $l=-1$   & & 0.192 & 0.096 & 0.057 \\
 \hline
 $l=0$ & 0.428    & 0.157 & 0.082 & 0.051\\
 \hline
 $l=+1$ & & 0.208 & 0.101 & 0.059\\
 \hline
\end{tabular}
\caption{Binding energies $|E_{nl}|$ (in eV) for hybrid exciton states labeled by the main $n=1,2,..$ and orbital $l=0,\pm1$ quantum numbers.}
\label{Tab1}
\end{center}
\end{table}

\section{IV. Excitons in motion }

For electrons and holes with a conventional quadratic spectrum, $C_\vec{pq}$ is independent of $\vec{q}$, which implies decoupling between relative motion and center-of-mass motion. However, this is not the case for hybrid excitons. To a linear order in $\vec{q}/p_{\Delta}$, the excitonic eigenvalue problem can be presented as follows 
\begin{equation}
\label{EigenvalueProblemFiniteQ}
\begin{split}
(\varepsilon^\mathrm{c}_{\vec{p}}-\varepsilon^\mathrm{v}_{\vec{p}}+\delta \varepsilon_{\vec{p}}) C_{\vec{p}\vec{q}} - \\ \sum_{\vec{p}'}  V_{\vec{p}-\vec{p}'}(\Lambda_{\vec{p} ,\vec{p}'}+\delta\Lambda_{\vec{p}\vec{p}'})C_{\vec{p}'\vec{q}}= E_\vec{q} C_{\vec{p}\vec{q}}.
\end{split}    
\end{equation}
The term $\delta \varepsilon_{\vec{p}}$ originates from the non-parabolicity of the spectrum for Dirac electrons and is given by 
\begin{equation}
\delta \varepsilon_{\vec{p}}=\vec{q}(\beta_\mathrm{c}\vec{\nabla}_\vec{p}\varepsilon_\vec{p}^\mathrm{c}+\beta_\mathrm{v}\vec{\nabla}_\vec{p}\varepsilon_\vec{p}^\mathrm{v})=\frac{\vec{p}\vec{q}}{M}\left(\frac{\Delta}{\varepsilon^\mathrm{c}_\vec{p}}-1\right).    
\end{equation}
The other term inside the integral
\begin{equation}
\delta\Lambda_{\vec{p}\vec{p}'}=\beta_\mathrm{c} \vec{q}(\nabla_\vec{p}+\nabla_{\vec{p}'}) \Lambda_{\vec{p}\vec{p}'}
\end{equation}
originates from the overlap of Dirac spinors, and its cumbersome expression can be derived from Eq.~(\ref{AngleFactor}) in a straightforward manner. 

The exciton wave function $C_{\vec{p}\vec{q}}$ for a finite momentum $\vec{q}$ can be decomposed in terms of the excitonic states $C_{\vec{p}}^{nl}$ at $\vec{q}=0$. The diagonal matrix elements of the eigenvalue problem, Eq.~(\ref{EigenvalueProblemFiniteQ}), are the excitonic energies $E_{nl}$. The off-diagonal matrix elements can be interpreted as coupling between excitonic states induced by motion. Here, we have two contributions 
\begin{equation}
\begin{split}
U_{nn'}^{ll'}=\sum_{\vec{p}} (C_\vec{p}^{nl})^* \delta \varepsilon_{\vec{p}} C_\vec{p}^{n'l'}, \\
W_{nn'}^{ll'}=\sum_{\vec{p} \vec{p}'} V_{\vec{p}-\vec{p}'} (C_\vec{p}^{nl})^* \delta \Lambda_{\vec{p} \vec{p}'} C_{\vec{p}'}^{n'l'}.\end{split}
\end{equation}
These matrix elements are obtained to a linear order in $q/p_\mathrm{\Delta}$ and are therefore nonzero only if $|l-l'|=1$, which can be viewed as the generalized dipole selection rule. 

\begin{figure}[t]
	\begin{center}
		\includegraphics[trim=0cm 0cm 0cm 0cm, clip, width=0.95\columnwidth]{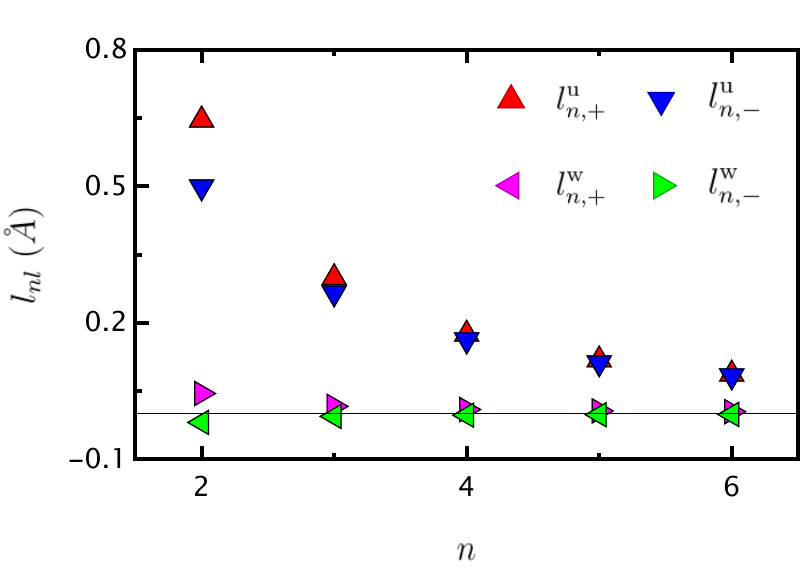}
		\caption{Lengths $l_{n,\pm}^\mathrm{u}$ and $l^\mathrm{w}_{n,\pm}$ describing the coupling between the ground and excited states induced by hybrid exciton motion. The lengths are defined in Eq.~(\ref{LengthDef}).}
		\label{Fig3}
	\end{center}
\end{figure}

We restrict our study to the ground hybrid exciton state, which is only relevant for the optical AVHE. The matrix elements for coupling between the ground and excited states can be parameterized in terms of the lengths $l^\mathrm{u}_{nl}$ and $l^\mathrm{w}_{nl}$~\footnote{According to their definition, the lengths $l^\mathrm{u}_{nl}$ and $l^\mathrm{w}_{nl}$ can be complex numbers, but an explicit evaluation shows that these lengths are real.}, which are defined as
\begin{equation}
\label{LengthDef}
\begin{split}
U_{n0}^{l0}=\beta_\mathrm{c} (E_{00}^\mathrm{X}-E_{nl}^\mathrm{X}) l^\mathrm{u}_{nl} q e^{-i l \phi_\vec{q}} \delta_{|l|,1},  \\ W_{n0}^{l0}=\beta_\mathrm{c} (E_{00}^\mathrm{X}-E_{nl}^\mathrm{X}) l^\mathrm{w}_{nl} q e^{-i l \phi_\vec{q}} \delta_{|l|,1}.
\end{split}
\end{equation}
The dependence of the lengths on the main quantum number is presented in Fig.~\ref{Fig3}. The coupling component due to the non-parabolicity of the electron dispersion $l_{nl}^{\mathrm{u}}$ is the strongest component. However, the asymmetry with respect to the orbital number $l$ is comparable for $l_{nl}^{\mathrm{u}}$ and $l_{nl}^{\mathrm{w}}$. For this reason, both mechanisms of inter-level coupling are essential.  

To a linear order in $q/p_{\mathrm{\Delta}}$, the coupling between the ground and exited states can be treated in a perturbative manner, and the wave function of the ground exciton state can be approximated as follows 
\begin{equation}
C_{\vec{p}\vec{q}}=C^{00}_{\vec{p}} + \sum_{n,l} \beta_\mathrm{c} l_{nl} q e^{-i l \phi_\vec{q}}  C^{nl}_{\vec{p}},\end{equation} where $l_{nl}=l_{nl}^\mathrm{u}+l_{nl}^\mathrm{w}$.  As illustrated in Fig.~\ref{Fig2}-b, the motion of the exciton couples its ground exciton state with the excited states.  The states become intertwined with the phase winding factors $e^{-i l \phi_\vec{q}}$, which are the usual hallmark for nontrivial geometries in coupled modes. Moreover, the contributions of excited states $l_{n\pm}$ with $l=\pm 1$ are not the same, which ensures that the phase windings with opposite chirality do not compensate each other.

\begin{figure}[t]
	\begin{center}
		\includegraphics[trim=0cm 0cm 0cm 0cm, clip, width=0.95\columnwidth]{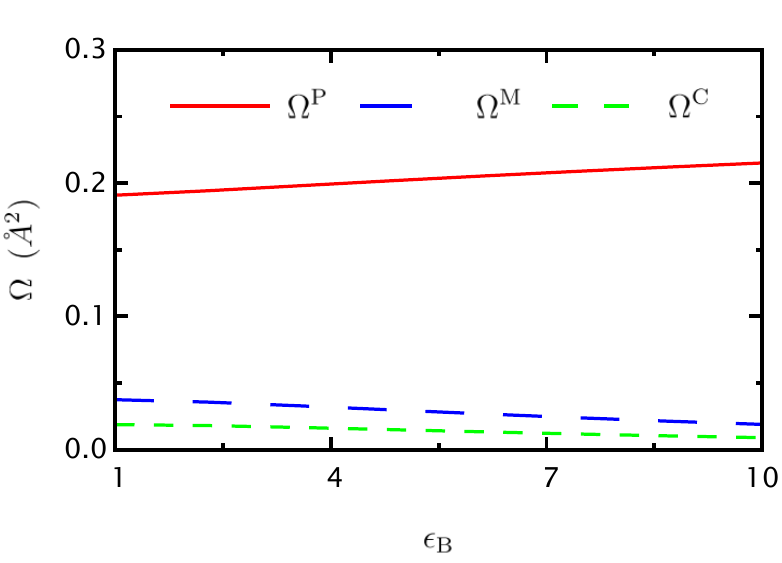}
		\caption{Dependence of the parent $\Omega^{\mathrm{P}}$, mixed $\Omega^\mathrm{M}$, and coupling induced $\Omega^\mathrm{C}$ contributions to the Berry curvature of hybrid excitons (see Eq.~(\ref{BerryCurvatureFull})) on the dielectric constant $\epsilon_\mathrm{B}$ of the substrate. In suspended samples ($\epsilon_{\mathrm{B}}=1$), additional terms, $\Omega^{\mathrm{M}}$ and $\Omega^\mathrm{C}$, account for almost one quarter of the total Berry curvature.}
		\label{Fig4}
	\end{center}
\end{figure}

\section{V. Berry curvature for excitons}

The nontrivial geometry of hybrid excitons is inherited from Dirac electrons. This geometry can be characterized by the Berry connection $\vec{A}^\mathrm{X}_\vec{q}=i \langle \vec{q}|\nabla_\vec{q}| \vec{q} \rangle_\mathrm{X}$ and curvature $\Omega^\mathrm{X}_{\vec{q}}= [{\bm \nabla_\vec{q}}\times {\bm A}^\mathrm{X}_{\vec{q}}]_\mathrm{z}$, which are defined in a manner similar to that for Dirac electrons. The Berry curvature $\Omega^\mathrm{X}_{\vec{q}}$ has three contributions of different physical origins, as identified in Ref.~\cite{BCQian}, which are given by 
\begin{equation}
\label{BerryCurvatureFull}
\begin{split}
&\Omega_{\vec{q}}^{\mathrm{P}} =\beta_\mathrm{c}^2\sum_{\vec p}|C_{{\vec p}{\vec q}}|^2 \Omega^{\mathrm{c}}_{\vec{p}_+}, \\
 &\Omega_{\vec{q}}^{\mathrm{C}} = \sum_{\vec{p} }\Big(\nabla_{\vec{q}}^{\mathrm{x}}C^*_{\vec{pq}} \nabla_{\vec{q}}^y C_{\vec{pq}} - \nabla_{\vec{q}}^{\mathrm{y}}C^*_{\vec{pq}} \nabla_{\vec{q}}^{\mathrm{x}}C_{\vec{pq}}\Big),\\
 & \Omega_{\vec{q}}^{\mathrm{M}}= \beta_\mathrm{c}\sum_{\vec p} \left[ \nabla_{\vec{q}}^{\mathrm{x}}|C_{\vec{p\vec{q}}}|^2 A^{\mathrm{c,y}}_{\vec{p}_+}- \nabla_{\vec{q}}^{\mathrm{y}}|C_{\vec{p\vec{q}}}|^2 A^{\mathrm{c,x}}_{\vec{p}_+}\right].
\end{split}
\end{equation}
The first term, $\Omega_{\vec{q}}^\mathrm{P}$, is the parent band contribution, which occurs because the Berry curvatures of excitons and electrons are directly related. The contribution, $\Omega_{\bm q}^\mathrm{C}$, reflects only intertwining between the ground and excited states induced by exciton motion. The mixed term, $\Omega_{\vec{q}}^\mathrm{M}$, originates from interference between the first two mechanisms. 

At small momenta $q\ll p_\mathrm{\Delta}$, the Berry curvature can be approximated by its value at $\vec{q}=0$ as 
\begin{equation}
\label{BerryCurvatureSimplified}
\begin{split}
\Omega^\mathrm{P}
=\beta_\mathrm{c}^2\sum_{\vec p}|C_{\vec{p}}^{10}|^2 \Omega^{\mathrm{c}}_{\vec{p}}, \quad \quad \Omega^\mathrm{C}=2 \beta_\mathrm{c}^2 \sum_{nl} l |l_{nl}|^2,
\\ \quad  \Omega^\mathrm{M}=2 \beta_\mathrm{c}^2 \sum_{nl} l_{nl} \mathcal{A}_{nl}^{\mathrm{c}},   \quad \mathcal{A}_{nl}^{\mathrm{c}}=\sum_\vec{p} C_{\vec{p}}^{10} A_\vec{p}^{\mathrm{c}} |C_{\vec{p}}^{nl}|.
\end{split}
\end{equation}
The terms $\Omega^\mathrm{C}$ and $\Omega^\mathrm{M}$ are presented as a function of the length $l_{nl}$, which describes the strength of mixing between the ground and excited excitonic states. Moreover, the term $\Omega^\mathrm{C}$ solely relies on the asymmetry of the lengths $l_{n,\pm}$ describing the coupling to excited states with with opposite quantum numbers $l=\pm1$. 

An explicit evaluation of these terms results in $\Omega^{\mathrm{P}}\approx0.199 \; \mathrm{\AA^2}$, $\Omega^\mathrm{C}\approx0.016 \; \mathrm{\AA^2}$, and $\Omega^{\mathrm{M}}\approx0.032 \; \mathrm{\AA^2}$. The parent band contribution clearly dominates, while the additional terms $\Omega^\mathrm{C}$ and $\Omega^{\mathrm{M}}$ account for $20\%$ of the total Berry curvature, $\Omega^{\mathrm{H}}\approx 0.247 \; \mathrm{\AA^2}$, for hybrid excitons. The additional contributions rely on the spectrum of excited exciton states, and their dependence on the dielectric constant $\epsilon_\mathrm{B}$ of the substrate below the $\hbox{MoS}_2$ monolayer is presented in Fig.~\ref{Fig4}. The total Berry curvature appears to be almost substrate-independent, but the additional terms $\Omega^\mathrm{C}$ and $\Omega^\mathrm{M}$ increase with decreasing $\epsilon_{\mathrm{B}}$ and reach $23\%$ in suspended samples.

The Berry curvature of hybrid excitons is smaller than that of $\Omega^\mathrm{B}\approx 4.1 \; \mathrm{\AA^2}$ for bright excitons by a factor of approximately $16$. This hierarchy arises because a substantial portion of the total momentum is carried by the heavy Schr\"{o}dinger hole, which is free of the nontrivial geometry. As a result, all contributions to the Berry curvature of the hybrid exciton listed in Eq.~(\ref{BerryCurvatureSimplified}) have a small prefactor $\beta_\mathrm{c}^2$. Moreover, the hierarchy can be tracked if we approximate the electronic Berry curvature given by Eq.~(\ref{ElectronBC}) by its value at the bottom of the conduction band $\Omega^\mathrm{c}=v_\mathrm{D}^2/2\Delta^2$. The Berry curvatures for hybrid and bright excitons are therefore approximated by their parent band contributions as $\Omega^\mathrm{H}\approx\beta_\mathrm{c}^2 \Omega^\mathrm{c}$ and $\Omega^\mathrm{B}=\Omega^\mathrm{c}/2$, respectively. Their ratio $\Omega^\mathrm{B}/\Omega^\mathrm{H}= 1/2 \beta_\mathrm{c}^2\approx 18$ matches reasonably well with explicit numerical calculations. In the next section, we will  argue that the small Berry curvature for hybrid excitons does not provide any disadvantage in the optical AVHE. 

\section{VI. Anomalous valley Hall effect}
The optical AVHE mediated by bright excitons has been reported in $\hbox{MoS}_2$ monolayers~\cite{EHE-2017} and heterostructures~\cite{EHE-2019,EHE-2020}. In this setup, bright excitons are optically excited at one side of a slab and propagate along the slab with valley-dependent transverse deflection. The density profiles of bright excitons from two valleys are tracked in real time via photoluminescence mapping. The anticipated presence of hybrid dark excitons (as well as intervalley dark excitons with energies comparable to those of bright excitons) has not yet been observed, but can be probed by a recently developed time-resolved photoemission spectroscopy technique~\cite{ExpImportant}. 

To describe the optical AVHE, we will follow Refs.~\cite{ExcitonDiffusion1,ExtrinsticTheory,PhononWind} and introduce a simplified model that ignores the coexistence of and conversion between bright and dark excitonic species. Instead, we assume that dark excitons are created by a source $S(\vec{r})$ and experience a deflected drift pushed by a temperature gradient or flow of phonons, which can be described by a (generalized) force $\vec{F}$. These excitons acquire a longitudinal drift velocity $\vec{u}_{\parallel}=\mu \vec{F}$, where $\mu=\tau_\mathrm{d}/M$ is the exciton mobility determined by the transport scattering time $\tau_{\mathrm{s}}$ due to disorder/phonons. These excitons also acquire a valley-dependent anomalous transverse velocity $\vec{u}_{\bot}=\kappa_\alpha [\vec{e}_\mathrm{z}\times \vec{F}]$. If we ignore the interplay between the nontrivial geometry and disorder and assume that the AVHE is intrinsic, $\kappa_\alpha$ is equal to the Berry curvature $\Omega_\alpha$ of hybrid excitons. The concentration of hybrid excitons $n_\alpha$ with valley index $\alpha$ satisfy the following coupled transport equations
\begin{equation}
\label{TransportEq}
\begin{split}
\frac{\partial n_\alpha}{\partial t}=D \Delta n_\alpha - \mu \vec{F}\cdot \nabla n_\alpha - \kappa_\alpha [\vec{F}\times \nabla n_\alpha]_z - \\  \frac{n_\alpha}{\tau} - \frac{n_\alpha-n_{\bar{\alpha}}}{\tau'}+S(\vec{r}).  
\end{split}
\end{equation}
Here, $D=\mu T$ is the exciton diffusion coefficient, which is related to the mobility $\mu$ by the Einstein relation $D=\mu T$. With the assistance of phonons, hybrid excitons can recombine with a lifetime of $\tau$. The intervalley relaxation is expected to be inefficient ($\tau\ll\tau'$) because electron hopping between two valleys requires not only a large momentum transfer but also a spin flip.

\begin{figure}[t]
	\begin{center}
		\includegraphics[trim=0.5cm 1cm 1.5cm 0cm, clip, width=0.95\columnwidth]{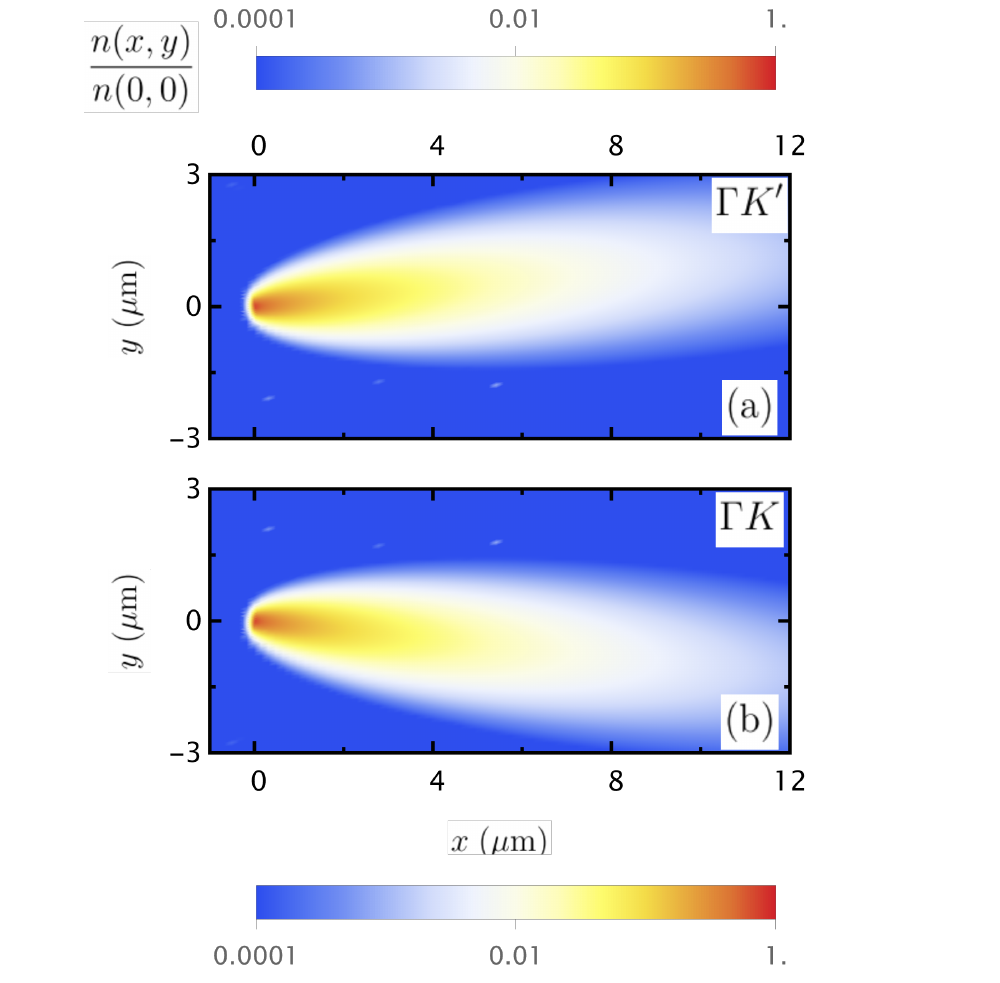}
		\caption{Density profile for  $\Gamma K'$ (a) and  $\Gamma K$ (b) hybrid excitons. Pushed by a temperature gradient or phonon flow, these excitons experience a valley-dependent deflected drift.}
		\label{Fig5}
	\end{center}
\end{figure}

The transport equation, Eq.~(\ref{TransportEq}), has multiple temperature- and disorder-dependent parameters, but its solution is generic, as presented in Fig.~\ref{Fig5} (the choice of parameters is discussed below). Excitons experience a deflected drift at the Hall angle $\tan\Theta_\alpha=\kappa_\alpha/\mu$ with respect to the force $\vec{F}$. The propagation distance and distribution broadening can be estimated by the drift $\xi_\mathrm{u}=u\tau$ and diffusion $\xi_\mathrm{d}=\sqrt{D \tau}$ lengths. Additionally, the shape of the exciton cloud is weakly dependent on the source profile unless its size $r_\mathrm{s}$ is comparable to the diffusion length, $\xi_\mathrm{d}$.   

For the case of bright excitons, the parameters of the transport equation, Eq.~(\ref{TransportEq}), have been estimated from experimental diffusion ~\cite{ExcitonDiffusion1,ExcitonDiffusion2,ExcitonDiffusion3} and AVHE~\cite{EHE-2017} data. Because the scattering time $\tau_{\mathrm{s}}$ for the short-range potential experienced by neutral excitons is inversely proportional to their mass $M$, the parameters for hybrid excitons can be obtained by rescaling. In particular, we obtain 
\begin{equation}
\begin{split}
\frac{\tan\Theta^\mathrm{H}}{\tan \Theta^\mathrm{B}}=\frac{\Omega^\mathrm{H} M_\mathrm{H}^2}{\Omega^\mathrm{B} M_\mathrm{B}^2}, \quad \frac{\xi_\mathrm{u}^\mathrm{H}}{\xi_\mathrm{u}^\mathrm{B}}=\frac{M_\mathrm{B}^2 \tau_\mathrm{H}}{M_\mathrm{H}^2 \tau_\mathrm{B}}, \quad \frac{\xi_\mathrm{d}^\mathrm{H}}{\xi_\mathrm{d}^\mathrm{B}}=\sqrt{\frac{\xi_\mathrm{u}^\mathrm{H}}{\xi_\mathrm{u}^\mathrm{B}}}. 
\end{split}
\end{equation}
Importantly, the ratio of the Hall angles is $\tan \Theta^\mathrm{H}/\tan \Theta^\mathrm{B}\approx 0.5$. The small Berry curvature for hybrid excitons and their large mass almost compensate each other. The lifetime of hybrid excitons has not yet been reported, but the lifetime for other dark species is usually more than an order of magnitude larger than the lifetime for bright excitons. Thus, the drift length $\xi^\mathrm{H}_\mathrm{u}$ can be considerably larger than $\xi^\mathrm{B}_\mathrm{u}$ , which favors the complete spatial separation of hybrid exciton clouds from different valleys. In addition, bright excitons suffer from efficient intervalley relaxation ($\tau'\ll\tau$), which originates from exchange interactions~\cite{IntervalleyExchange1,IntervalleyExchange2,IntervalleyExchange3,IntervalleyExchange4}. 

The transport scattering time can be estimated as $\tau_\mathrm{s}^\mathrm{B}\approx30\;\hbox{fs}$ and $\tau_\mathrm{s}^\mathrm{H}\approx10\;\hbox{fs}$, which results in Hall angles of $\tan \Theta^\mathrm{B}\approx0.025$ and $\tan \Theta^\mathrm{H}\approx0.012$, respectively. The estimated Hall angle for bright excitons is moderate and observable, but is approximately one tenth of that reported for experiments, $\tan \Theta^\mathrm{B}\approx0.2$~\cite{EHE-2017}. This result clearly demonstrates that the observed AVHE cannot solely rely on an intrinsic mechanism. The microscopic theory of an extrinsic contribution~\cite{ExtrinsticTheory} arising from the interplay between nontrivial geometry and disorder is essential, but is beyond the scope of the present work. 

For numerical simulations of the transport equation, Eq.~(\ref{TransportEq}), we assumed that the source $S(\vec{r})$ has a Gaussian profile with size $r_\mathrm{s}=60~\hbox{nm}$ and used the following set of parameters: $\tau_\mathrm{s}=10\;\hbox{fs}$, $D=0.03~\hbox{sm}^2/\hbox{s}$, and $u=0.1\;\mu\hbox{m}/\hbox{ns}$. The lifetime $\tau\approx 30~\;\hbox{ns}$ is chosen to be $30$ times larger than and the Hall angle $\tan\Theta=0.1$ to be one half of experimentally reported values for bright excitons~\cite{ExcitonDiffusion1,EHE-2017,ExtrinsticTheory}. We have also neglected the intervalley relaxation. The solution of the transport equation is presented in Fig.~\ref{Fig5}, which shows the valley-dependent deflected drift of the hybrid excitons and their spatial separation. For the parameters given above, the drift and diffusion lengths are given by $\xi_\mathrm{u}\approx 3\; \mu\hbox{m}$ and $\xi_\mathrm{d}\approx 0.32\; \mu\hbox{m}$ and roughly determine the parameters of the exciton clouds.   

\section{VII. Discussion}
The rich exciton landscape in $\hbox{MoS}_2$ includes a  variety of dark excitonic states. The energies of most of these states are above the energy of bright excitons, and therefore, they are of little importance. However, this is not the case for intervalley dark excitons (formed by an electron in $K$ valley and a hole in $K'$ one or vice versa), which also play an important role in the thermalization of photoexcited bright excitons. These intervalley dark excitons are geometrically trivial (all contributions to the Berry curvature are zero), and their drift does not experience any transverse deflection. 

The additional contributions to the Berry curvature ($\Omega^\mathrm{C}$ and $\Omega^\mathrm{M}$) due to the coupling between  
the center-of-mass motion and relative motion for the electron and hole are unique feature of hybrid excitons. These contributions rely on the intricate interplay between the exciton motion-induced coupling among the ground and excited exciton states and their asymmetry. The additional contributions vanish for both bright and intervalley dark excitons in TMD monolayers. In the former case, the spectrum exhibits asymmetry, but there is no exciton motion-induced coupling between the ground and excited states~\footnote{The absence of motion-induced coupling relies on the electron-hole asymmetry observed for the massive Dirac model but not for lattice tight-binding models describing electronic structure of $\hbox{MoS}_2$ monolayers. However, for the low-energy conduction and valence band states, the symmetry violations are negligibly small.}. For the latter case, coupling is present, but the spectrum is symmetric. 

Due to the presence of a valley degree of freedom for Dirac electrons and the spin of one for Schr\"{o}dinger holes, the hybrid excitonic states have quartic degeneracy~\footnote{Generally, the degeneracy can be lifted by the spin-orbit interactions for the Schrodinger holes. However, the the extensive DFT calculations~\cite{Korm_nyos_2015} demonstrate that the latter are negligibly small}. The doublet states are both momentum- and spin-forbidden, but the second pair is only momentum-forbidden. The states have the same Berry curvature, but their lifetimes are anticipated to differ greatly.  

The conversion between hybrid and bright excitons requires a momentum transfer of $\vec{K}$ or $\vec{K}'$. This transfer can be achieved via valley phonons that have been recently identified as chiral in TMD monolayers~\cite{ChiralPhononTheory}. The chiral phonons carry orbital momenta that are opposite for two valleys. As recently demonstrated in ultraclean $\hbox{WSe}_2$, intervalley excitons recombine with the assistance of chiral phonons, and the corresponding selection rules are similar to those of bright excitons~\cite{ChiralPhononExp1,ChiralPhononExp2,ChiralPhononExp3,ChiralPhononExp4}. In a similar fashion, we expect chiral phonons to provide valley momentum to the Schr\"{o}dinger hole to assist in the recombination of hybrid excitons. 

To conclude, we have uncovered the nontrivial geometry of hybrid excitons in $\hbox{MoS}_2$ monolayers. A unique feature of these excitons is an additional contribution to the Berry curvature due to coupling between the center-of-mass motion and relative motion for the electron and hole.  Our analysis of the intrinsic AVHE has demonstrated that the Hall angle of hybrid excitons is approximately one half of that for bright excitons, while their expected long lifetime favors an extended propagation distance of hybrid excitons.  Our predictions can be tested by state-of-the-art approaches, and it is anticipated that these findings will drive further experiments with $\hbox{MoS}_2$.   

\section{Acknowledgements}

We acknowledge fruitful discussions
with Michael Fuhrer, Shao-Yu Chen, and Jeff Devis and support from the Australian Research Council Centre of Excellence in Future Low-Energy Electronics Technologies. 
\bibliography{ref-exciton}

\end{document}